\documentclass[12pt]{iopart}
\usepackage{iopams}  
\usepackage{graphicx}
\usepackage{bm}
\begin{document}

\title[Spatio-temporal chaotic synchronization as an absorbing phase transition...]
{Synchronization of spatio-temporal chaos as an absorbing phase transition: a study in $2+1$ dimensions}

\author{Francesco Ginelli$^1$, Massimo  Cencini$^2$  and Alessandro Torcini$^3$}

\address{$^1$ Institut des Syst\'emes Complexes Paris \^Ile-de-France,
  57-59 rue Lhomond, 75005 Paris, France and CEA - Service de Physique
  de l'\'Etat Condens\'e, Centre d'\'Etudes de Saclay, 91191
  Gif-sur-Yvette, France}

\address{$^2$ INFM-CNR, SMC Dipartimento di Fisica, Universit\`a di
  Roma ``La Sapienza'', p.zle A.\ Moro 2, 00185 Roma, Italy and
  Istituto dei Sistemi Complessi - CNR, via dei Taurini 19, 00185
  Roma, Italy}

\address{$^3$ Istituto dei Sistemi Complessi - CNR, via Madonna
  del Piano 10, 50019 Sesto Fiorentino, Italy and INFN - Sezione
  di Firenze and CSDC, via Sansone 1, 50019 Sesto Fiorentino, Italy}

\begin{abstract}
  The synchronization transition between two coupled replicas of
  spatio-temporal chaotic systems in 2+1 dimensions is studied as a
  phase transition into an absorbing state --- the synchronized state.
  Confirming the scenario drawn in 1+1 dimensional systems, the
  transition is found to belong to two different universality classes
  --- Multiplicative Noise (MN) and Directed Percolation (DP) ---
  depending on the linear or nonlinear character of damage spreading
  occurring in the coupled systems.  By comparing coupled map lattice
  with two different stochastic models, accurate numerical estimates
  for MN in $2+1$ dimensions are obtained. Finally, aiming to pave the
  way for future experimental studies, slightly non-identical replicas
  have been considered.  It is shown that the presence of small
  differences between the dynamics of the two replicas acts as an external field in the
  context of absorbing phase transitions, and can be characterized in
  terms of a suitable critical exponent.
\end{abstract}


\submitto{Journal of Statistical Mechanics: Theory and Experiments}
\maketitle

\section{Introduction}
\label{sec:intro}
Synchronization of low dimensional chaotic oscillators, which have
been intensively investigated in the last two decades \cite{PikoBook},
can be essentially regarded as a bifurcation from an uncorrelated to
an entrained state.  When single low dimensional oscillators are
replaced by spatially extended systems exhibiting spatio-temporal
chaos, however, synchronization becomes a genuine, fluctuation driven,
phase transition which separates the uncorrelated and the completely
synchronized states \cite{baroni,PA02}.  The synchronization
transition (ST) between two identical coupled replicas (starting from
different, generic initial conditions) of the same spatio-temporal
chaotic dynamics is thus a non-equilibrium critical phenomenon
originating from deterministic dynamics. Chaotic fluctuations on the
synchronized trajectory play the role of intrinsic stochastic terms,
leading to diverging fluctuations as the critical point is approached
\cite{PPK}.

Numerical studies in one dimensional dissipative extended systems have
shown that the ST is essentially a phase transition into an absorbing
state~\cite{haye}, i.e. the completely synchronized state, in which
the two replicas of the extended system evolve on the same chaotic
trajectory~\cite{baroni,PA02}, similarly to low dimensional chaotic
systems \cite{PG91}. Noticeably, the observed continuous STs can be
distinguished in two universality classes depending on the
spatio-temporal propagation properties of the system close to
criticality~\cite{baroni}. Consider a localized and finite
perturbation (or {\it synchronization error}) to one of two otherwise
identical (and thus synchronized) coupled replicas. At low coupling it
will spread with a well defined propagation velocity $v_F$
\cite{Ginelli2003}, implying a desynchronization of the two
replicas. By definition, ST takes place at the critical coupling value
for which $v_F$ vanishes, i.e. the synchronization error does not
propagate anymore in space and time.  On the other hand, the evolution
of ``infinitesimal'' synchronization errors is well described by the
linearized dynamics and can be characterized in terms of the asymptotic
exponential growth rate, i.e.  the so-called {\it transverse Lyapunov
  exponent} $\lambda_T$ \cite{PikoBook, PG91}: a positive
(resp. negative) $\lambda_T$ implies exponential growth
(resp. contraction) of infinitesimal perturbations.  Whenever the
local dynamics is sufficiently smooth, i.e. the linearization captures
the essential features of the full dynamics, $\lambda_T$ vanishes
together with $v_F$ at the critical coupling for the
synchronization. When this happens ST belongs to the Multiplicative
Noise (MN)\footnote{Or, to be more precise, to the so called MN1
  class, see for instance Ref. \cite{Munoz2003}. In this paper MN will
  always indicate the MN1 class.} universality class
\cite{munoz_review}.  When strong nonlinear effects make finite
amplitude perturbations more unstable than inﬁnitesimal ones
\cite{cencini_torcini}, the transverse Lyapunov exponent vanishes
before the error propagation velocity, so that at the ST transition
one has $\lambda_T<0$ and $v_F=0$. In this second case, which is
closely related to the {\it Stable Chaos} phenomenon \cite{SC}, ST
belongs to the Directed Percolation (DP) universality class
\cite{haye}.  In the framework of one dimensional coupled map lattices
(CMLs) -- a common prototype of spatio-temporal chaotic systems -- the
critical behavior belongs to the DP universality class for
(almost-)discontinuous local maps, and to the MN class for smoother
maps~\cite{baroni,PA02}. Remarkably, in the context of CMLs models, it
is possible to pass from DP to MN by varying a unique parameter, which
controls the strength of nonlinear instabilities of the local
map~\cite{rechtman}.

Analytical arguments (coarse-graining techniques combined with
linearization or finite-size analysis) show that these two STs can be
mapped into the Langevin equations describing the MN or DP
universality classes, respectively \cite{PA02,Ginelli2003}.  It is
worth noticing that these two universality classes have been
originally identified in completely different contexts, such as
epidemics spreading in reaction diffusion dynamics (DP) or the
depinning of a fluctuating Kardar-Parisi-Zhang (KPZ) \cite{KPZ}
interface (with a {\it negative} nonlinear term) from an underlying
substrate (MN)\footnote{A second universality class, called MN2, can
  be defined starting from a KPZ with a {\it positive} nonlinear term.
  However, it is completely unrelated to synchronization
  problems.}. Interestingly, while it is not possible to define a
simple reaction diffusion system with a Markov dynamics belonging to
the MN class, DP critical depinning can be observed when a short
ranged attraction force is imposed between a KPZ interface and the
underlying substrate \cite{Ginelli2003b}. Therefore, ST can be
described in terms of a single Langevin
equation~\cite{munoz}. Although it is yet unclear how the empirical
Langevin equation for ST may eventually flow -- as a control parameter
is varied from the MN to the DP region -- to the same DP fixed point
as the Langevin equation for Reggeon field theory, these results
strongly suggest that both universality classes may be described
within a unique field-theoretic framework.

In the last few years, ST has been object of intensive numerical
investigations \cite{baroni, PA02,Ginelli2003, rechtman, droz, cecconi, gade,szendro09}; the analysis has
also been extended to stochastic coupling \cite{baroni}, 
cellular automata \cite{Bagnoli99, Morvan},
systems with
long-range interactions \cite{Tessone2007,Cencini2008} and delayed
dynamical systems~\cite{mohanty,szendro}, exploiting the analogy
between the latter and spatially extended dynamics. So far, however,
ST critical behavior has been investigated in 1+1 dimensions only.  In
this respect, it should be also remarked that while DP critical
exponents are known with great accuracy in any dimension, and MN
exponents have been accurately measured in 1+1 dimensions
\cite{Ginelli2005}, the only known estimates of the MN critical
indexes in 2+1 dimensions comes from direct numerical simulations of
the Langevin dynamics~\cite{MN2}. Moreover, it is worth recalling
that, for $d\leq 2$ the MN critical behavior is completely governed by
a {\it strong coupling fixed point} \cite{Munoz96}, so that no field
theoretical estimates of critical exponents are currently available.

The above situation is particularly unsatisfactory, especially
considering that two dimensional spatio-temporal chaotic systems --
such as chemical turbulence in quasi-two dimensional reactions
\cite{Kuramoto, exper} or turbulent nematic liquid crystals
\cite{Chate} -- are the most promising ones for studying ST in
experiments.  In this perspective, it is worth stressing that no
experimental realization of the MN class has been so far realized, and
it is only recently that the DP critical exponents have been measured
in experiments: in 1+1~\cite{tedeschi} and 2+1
dimensions~\cite{Chate}.  Therefore, besides the potential interest of
experimentally realizing synchronization of chaotic extended systems,
such an experiment would be the perfect ground for testing MN critical
behavior in a physical framework.

A major difficulty in devising a synchronization experiment is the
assemblage of two perfectly identical replicas of the same system. In
practice, it would be unavoidable to experience very small
differences, e.g., in the physical parameters entering the dynamics of
the two systems.  It is therefore of practical interest to quantify
the influence of a small mismatch in the copies of the two
systems. Another difficulty could lie in producing a suitable local
coupling between the two replicas.  However, this can be also realized
through a stochastic forcing, for instance exposing two excitable
chemical samples to the same random illumination (see, e.g.,
\cite{exper}), or by applying the same random external voltage to two
replicas of the intermittent electrohydrodynamic convection regimes of
Ref.\cite{Chate}. As far as the critical properties of ST are
concerned, deterministic or stochastic couplings are expected to share
the same properties \cite{baroni,PA02}.

This paper focuses on the synchronization transition in two spatial
dimensions, within the framework of CMLs, with a twofold scope. First,
we aim at verifying whether the scenario drawn in one dimensional
systems for ST applies also to two dimensions. This requires accurate
estimates of the MN critical exponents. For such reason we also
investigate two stochastic models which are expected to belong to MN
universality class.  Second, mimicking what could happen in an
experiment, we study ST in the presence of a small parameter mismatch
between the two replicas.

The paper is organized as follows. In Sect.~\ref{sec:uno}, after
presenting the CMLs models and recalling the definition and basic
tools for measuring the critical exponents, we discuss the results of
accurate numerical simulations for two different classes of maps,
which in 1+1 dimensions have been proved to belong to the DP and MN
universality classes. Sect.~\ref{sec:due} focuses on accurate estimation of
critical exponents for MN class in $2+1$ dimensions by means of two
stochastic models. The section ends with a critical comparison between
the computed exponents for the MN universality class and some scaling
relations which have been put forward by previous studies.  In
Sect.\ref{sec:tre}, mimicking what would typically be an experimental
setting, we reconsider ST in the presence of a parameter mismatch and,
accordingly, we introduce a new critical exponent for its
characterization.  Sect.~\ref{sec:quattro} is devoted to
final remarks and discussions.

\section{\label{sec:uno} Synchronization transition in two spatial
  dimensions}
In the following we investigate the dynamics of two coupled replicas
of two dimensional CMLs, defined on a $L\times L$ square lattice with periodic
boundary conditions, evolving according to the dynamics:
\begin{eqnarray}
u_{t+1}(x,y) &=& (1-\gamma) F(\tilde{u}_t(x,y))+\gamma F(\tilde{v}_t(x,y))\nonumber\\
v_{t+1}(x,y) &=& (1-\gamma) F(\tilde{v}_t(x,y))+\gamma F(\tilde{u}_t(x,y))
\label{CML}
\end{eqnarray}
with $x,y=1,2,\ldots,L$.  The variable
\begin{eqnarray}
\tilde{z}_t(x,y)&=&\nabla^2_\varepsilon z_t(x,y) \equiv (1-\varepsilon) z_t(x,y) \\
&+&\frac{\varepsilon}{4}[z_t(x+1,y)+z_t(x-1,y)+
z_t(x,y+1)+z_t(x,y-1)] \,,\nonumber
\label{diff}
\end{eqnarray}
(with ${z}={u},{v}$) represents the nearest-neighbour diffusive
coupling within each replica. Through this work, we set the diffusive
coupling constant $\varepsilon$ equal to the democratic value
$\varepsilon=4/5$, that gives same weight to all neighbors.  As from
studies in the one dimensional version of Eq.~(\ref{CML}), the DP or
MN character of the ST relates to the functional form of the local map
$F(u)$ \cite{baroni,PA02}. In particular, 
we have that if $F(u)$ is continuous (such as the logistic
or the tent map) the transition is in the MN universality class.
Conversely, for discontinuous (or quasi-discontinuous \cite{PT94})
maps, such as the generalized shift map 
\begin{equation}
F(u)=a u \quad \mathrm{mod}\;1\,,
\label{Bernoulli}
\end{equation}
the ST belongs to the DP-class. 
The strength of the ``transverse'' coupling between the replicas is
controlled by $\gamma$: for $\gamma=0$ they are completely
uncorrelated, while setting $\gamma=1/2$ induces trivial complete
synchronization in one time step, i.e. $u_t(x,y)=v_t(x,y)$ for any
$t\geq 1$ and for all $x,y$. Nontrivial ST, if present, is expected
for a critical coupling value $\gamma_c\in\, ]0:1/2[$ at which the
synchronized state, $u_t(x,y)=v_t(x,y)$, becomes stable (or, at least,
marginally stable). For any $\gamma \geq \gamma_c$, the two replicas
(starting from different, generic initial conditions) converge toward
the same spatio-temporal chaotic trajectory. In other terms, the
synchronization error field $w_t(x,y)=|u_t(x,y)-v_t(x,y)|$ tends
towards zero for any $x,y$. The linear  stability
properties of the synchronized state are ruled by the transverse
Lyapunov exponent $\lambda_T$. For $\gamma\geq\gamma_c$, $\lambda_T$
can be directly computed from the maximum Lyapunov exponent $\lambda$
of an uncoupled replica according to the relation~\cite{PG91}
\begin{equation}
  \lambda_T=\lambda+\ln (1-2\gamma)\,.
\label{lambdaT}
\end{equation}
The request $\lambda_T=0$ thus determines the coupling
$\bar{\gamma}=(1-\exp(-\lambda))/2$, which coincides with the critical
coupling for ST, i.e. $\bar{\gamma}=\gamma_c$, for maps belonging to
the MN class. However, this result, based on the linear analysis,
does not hold for systems belonging to the DP universality class.

The suitable order parameter to characterize ST is the
spatial average of the synchronization error
$w_t(x,y)=|u_t(x,y)-v_t(x,y)|$, i.e.
\begin{equation}
\rho(t)= \frac{1}{L^2} \sum_{x,y=1}^{L} w_t(x,y)\,.
\label{OP}
\end{equation}
Note that $\rho(t)=0$ if and only if $w_t(x,y)=0$ for all $x,y$. If
two replicas are identical (i.e. synchronized) at time $t_0$, they
will remain so at all times $t\geq t_0$, implying that the
synchronized state is {\it absorbing}, i.e. the dynamics cannot 
escape from it. 

Interestingly, the MN and DP absorbing states differ in their measure
\cite{munoz98} which, for any finite system size $L$, is vanishing or
finite, respectively. As a consequence any finite system belonging to
the DP class falls into the absorbing state in a finite time.  In
CMLs, whose state variables are continuous, the synchronized state
$w_t(x,y)=0$ can only be reached asymptotically in time (barring
computer round-offs), apparently at odd with the nature of the DP
absorbing state. In Ref.~\cite{Ginelli2003}, however, it has been
shown that whenever the ST transition takes place at $v_F=0$ and
$\lambda_T<0$ (the DP case), all perturbations smaller than a certain
{\it finite} (but vanishing for $L\to 0$) threshold are exponentially
contracted towards zero, thus defining an effective finite measure for
the absorbing state.

We now consider the scaling theory for phase transitions into an
absorbing state \cite{haye}, which allows for defining the {\it
  universal critical exponents}, only depending on properties such as
system symmetries and spatial dimensions, in terms of which we can
characterize the ST.  In the thermodynamic limit and at
$\gamma=\gamma_c$, one expects the order parameter to follow the
asymptotic power law decay
\begin{equation}
\rho(t)\sim t^{-\theta}\,,
\label{theta}
\end{equation}
where $\theta$ is a critical exponent.  Close to the critical point,
but slightly within the unsynchronized phase, i.e.  for
$\gamma<\gamma_c$ and $\gamma_c-\gamma\ll 1$, the asymptotic value of
the order parameter
$\rho_\infty \equiv \lim_{T \to \infty} \frac{1}{T}\sum_{t=1}^T \rho(t)$
is characterized by the scaling relation 
\begin{equation}
\rho_\infty \sim (\gamma_c-\gamma)^\beta\,,
\label{beta}
\end{equation}
which defines a second critical exponent.  Out of equilibrium critical
phenomena are also characterized by the divergence of space and time
correlations close to the critical point, in particular one expects
\begin{equation}
\xi_\parallel \sim |\gamma-\gamma_c|^{\nu_\parallel}\,,\qquad\quad
\xi_\perp \sim |\gamma-\gamma_c|^{\nu_\perp} 
\label{xi}
\end{equation}
where $\xi_\parallel$ (resp. $\xi_\perp $) is the temporal
(resp. spatial) correlation length, where for simplicity we considered
the system to be spatially isotropic. It can be shown \cite{haye} that
only three exponents are independent, and in particular that the
following relation holds
\begin{equation}
\theta = \beta \nu_\parallel\,.
\end{equation}
Furthermore, the ratio $z=\nu_\parallel/\nu_\perp$,  i.e. the so-called
dynamical exponent, determines the relation between time and space
correlations. The dynamical exponent can be measured exploiting the
 finite-size scaling relation 
\begin{equation}
\rho(t) = L^{-\theta z} f(t/L^z)\,,
\label{z}
\end{equation}
which holds at the
critical point ($f$ being a universal scaling function).

\begin{figure}[t!]
\centering
\includegraphics[clip=true,keepaspectratio,width=1\textwidth]{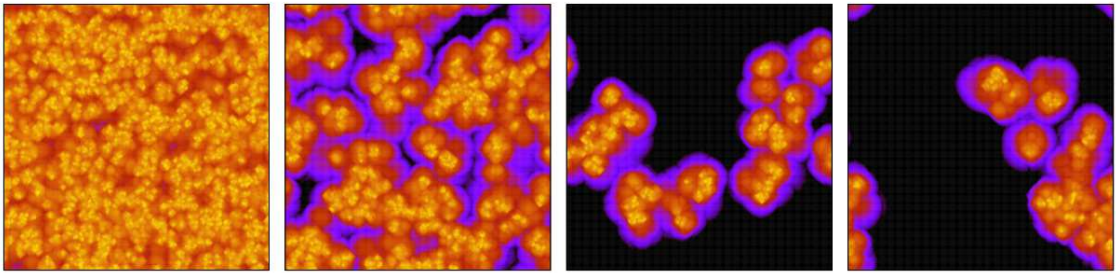}
\caption{Snapshots of the synchronization error field $w_t(x,y)$ at
  four successive times (increasing from left to right) for a
  $2d$-lattice of $256\times 256$ coupled Bernoulli shift maps.  The
  two replicas are transversely coupled with $\gamma=0.32824$,
  slightly above the critical point. Color code goes from light toward
  black indicating decreasing intensity of the field (in logarithmic
  scale).}
\label{fig:1}
\end{figure}

\subsection{Discontinuous Maps and Directed Percolation universality class}

We start analyzing the synchronization transition of two different
replicas of the same CML (\ref{CML}) whose local dynamics is given by
a discontinuous map. In particular, we consider the Bernoulli map
(\ref{Bernoulli}) with $a=2$.  Fig.~\ref{fig:1} displays the typical
spatial structure of the difference field $w_t(x,y)$, as obtained
iterating two replicas starting from (independent) random
initial conditions, at successive times, for $\gamma$ slightly larger
than the critical coupling value.  The figure reveals typical DP
patterns in proximity of the synchronization threshold.  In order to
make quantitative such observation, we need to measure the critical
exponents. To this aim, we preliminarily determined the critical point
by a careful finite-size analysis of the order parameter time
decay. In particular, considering systems' sizes up to $L=4096$, we
obtained $\gamma_c=0.32817(2)$.  Being the maximum Lyapunov exponent
of the single CML $\lambda=\ln 2$, from Eq.~(\ref{lambdaT}) one can
derive $\lambda(\gamma_c) = -0.3749$.  Thus the ST takes place at a
definitely negative Lyapunov exponent and the synchronized state is
truly absorbing~\cite{Ginelli2003}.

\begin{figure}[t!]
\centering
\includegraphics[clip=true,keepaspectratio,width=.9\textwidth]{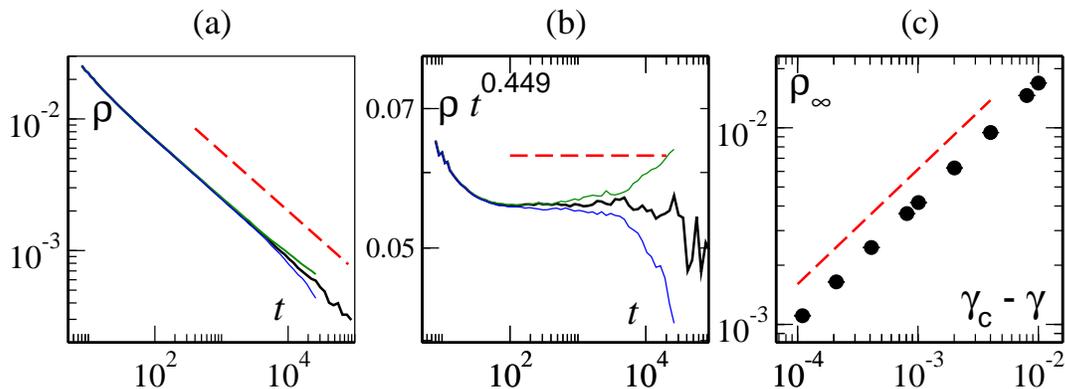}
\caption{Critical behaviors of ST for CMLs of Bernoulli maps, data
  refer to a $1024\times 1024$ lattice.  (a) Order parameter as a
  function of time at criticality $\gamma_c=0.32817$ (black line) and
  slightly below ($\gamma=0.32815$) and above ($\gamma=0.32819$) the
  critical point (coloured thin lines). Data are averaged over $200$
  initial conditions. The dashed red line marks the $t^{-\theta}$
  behavior with $\theta=0.449$.  (b) The compensated order parameter
  $\rho(t) t^{\theta}$ reveals a satisfactory plateau over about two
  decades.  (c) Asymptotic order parameter $\rho_\infty$ as a function
  of the distance from the critical point. The dashed red line shows
  the best fit with $\beta=0.584(9)$. For each $\gamma$, $\rho_\infty$
  is obtained averaging over $100$ initial conditions.}
\label{fig:2}
\end{figure}

We evaluate the exponent $\theta$ at criticality by averaging the
instantaneous order parameter over about $200$ independent initial
conditions in systems of size $L=1024$, obtaining a convincing
straight line in a doubly logarithmic plot (Fig.~\ref{fig:2}a).  The
best fit provides $\theta=0.449(4)$, which is perfectly compatible
with the best DP numerical estimates in $2+1$ dimensions, that is
$\theta_{DP}=0.451(6)$ \cite{haye}.  The quality of such a scaling law
is tested by multiplying the order parameter $\rho(t)$ by
$t^{\theta}$, as shown in in Fig.~\ref{fig:2}b we obtain an almost
two decades long plateau.

We next compute $\rho_\infty$ for several $\gamma$ below the critical
point at $L=1024$. By averaging over $\approx 100$ different initial
conditions we estimate $\beta=0.584(9)$ (Fig.~\ref{fig:2}c), to be
compared with the DP estimate $\beta_{DP}=0.584(4)$.  The larger error
is mainly due to the uncertainty on the location of the critical
point.

\begin{figure}[t!]
\centering
\includegraphics[clip=true,keepaspectratio,width=.7\textwidth]{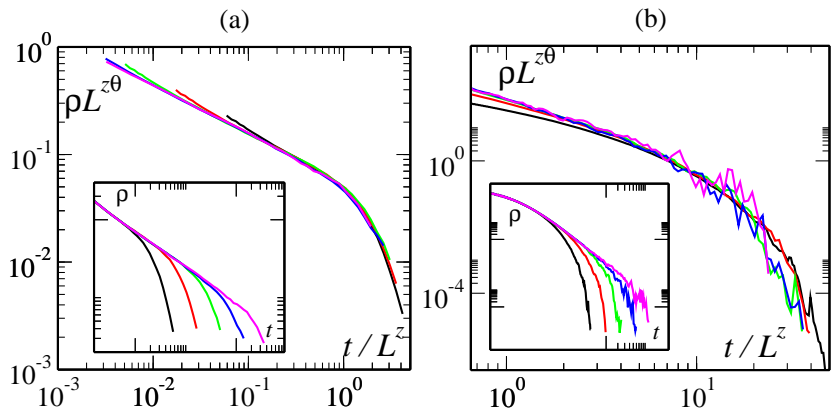}
\caption{The finite size scaling collapse for ST in CMLs 
  according to Eq.~(\ref{z}) is employed to estimate the exponent
  $z$. (a) Collapse for CMLs of Bernoulli maps obtained using
  $\theta=0.449$ and $z=1.77$.  Inset: non rescaled data for $L=16,
  32, 64, 128, 256$ (from left to right).  (b) Collapse for skewed
  tent maps (see Sec. \ref{sec:unob}) obtained using $\theta=1.81$,
  $z=1.55$.  Inset as in (a). Data have been averaged from $10^2$ up
  to $10^4$ realizations depending on the system size}
\label{fig:3}
\end{figure}

Finally, we determine the dynamical exponent $z$ through the
finite-size scaling relation (\ref{z}). Our best data collapse (shown in
Fig.~\ref{fig:3}a) suggests  $z=1.77(3)$. All together, the three
critical exponents $\theta$, $\beta$ and $z$
completely identify the DP universality class.

In Table~\ref{table} the results are summarized together with the
exponent for DP in $d=2$ as obtained from the best known numerical
estimates reported in Ref.~\cite{haye}. The agreement is very good and
we can safely affirm that the ST of map with discontinuities in $d=2$
is in the DP universality class, confirming the one dimensional
findings.

\begin{table}[hbt]
\centering
\begin{tabular}{||c|c|c|c||} \hline 
      & $\theta$ & $\beta$ & $z$ \\\hline \hline
Bernoulli  &   0.449(4)  &   0.584(9) &  1.77(3) \\
DP \cite{haye}  & 0.451(6) &  0.584(4)   &  1.76(3) \\
\hline
\end{tabular}
\caption{Critical exponents for coupled
  Bernoulli maps in $2+1$ dimensions together with
  the best estimations of the critical indexes for DP in $d=2$.
  For DP $\theta$ and $z$ have been derived by employing
  $\nu_\parallel=1.295(8)$ and $\nu_\perp=0.733(6)$ \cite{haye}.}
\label{table} 
\end{table}

\subsection{\label{sec:unob} Continuous Maps and Multiplicative Noise universality class}
\begin{figure}[t!]
\centering
\includegraphics[clip=true,keepaspectratio,width=1\textwidth]{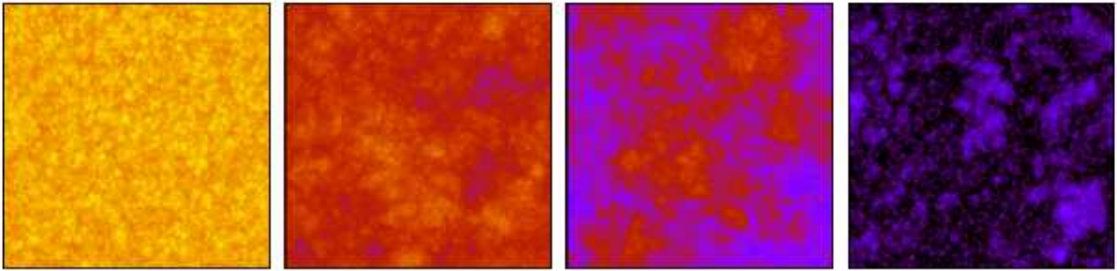}
\caption{Same as Fig.\ref{fig:1} for coupled skewed tent maps. The
  transversal coupling is $\gamma=0.132$, slightly above the
  critical point. The color coding and the scale are the same as
  Fig.\ref{fig:1}.}
\label{fig:snapMN}
\end{figure}

We now study the synchronization transition in CMLs with continuous
local maps. In particular, we consider the system (\ref{CML}) with
local dynamics given by the skewed tent map on the unit interval,
namely
\begin{equation}
F(x) = \left\{
\begin{array}{ll}
a x &\mbox{if}\,\, x \leq 1/a \\
a (x-1)/(1-a) &\mbox{if}\,\, x > 1/a \\
\end{array}
\right.\,
\label{skewed}
\end{equation}
where we set $a=2.2$. The skewed tent map is the simplest generic
continuous map; the skewness ensures the fluctuation of the
multipliers (first derivatives of the map) in tangent space, which is
the generic behavior \cite{PG91}.  Similarly to Fig.~\ref{fig:1},
Fig.~\ref{fig:snapMN} illustrates the spatio-temporal evolution of the
synchronization error field $w_t(x,y)$ for $\gamma$ slightly larger
than the critical coupling value.  Already at first glance, comparing
the two figures one can argue that the two STs should belong to
different universality classes. It is also apparent the different
nature of the absorbing state (in black).

To make quantitative the above statement, as for the previous case, we
first identified the critical coupling $\gamma_c$ at which ST takes
place.  

From the time decay of the order parameter $\rho(t)$ we estimated
$\gamma_c=0.131760(5)$. This value appears to be compatible with the
requirement that the transverse Lyapunov exponent $\lambda_T$ vanishes
at the MN synchronization transition.  Indeed, from an independent
numerical simulation of a single replica, we found that the largest
Lyapunov exponent is $\lambda=0.30588(6)$, which through
Eq.~(\ref{lambdaT}) yields $\lambda_T=0.00001(6)$.  Thus the ST now
takes place at a zero transverse Lyapunov exponent.  It should be
mentioned that to minimize finite size effects, which are rather
severe in this case (as discussed in the following) we have considered
lattice up to size L=8192.

\begin{figure}[t!]
\centering
\includegraphics[clip=true,keepaspectratio,width=.9\textwidth]{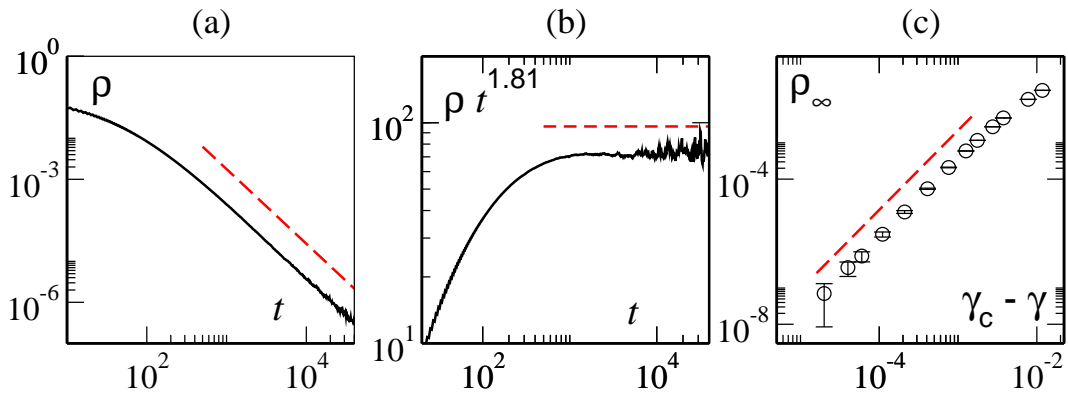}
\caption{Critical behavior of ST for CML's of skewed tent maps
  (\ref{skewed}) with $a=2.2$. (a) Power law decay of the order
  parameter at $\gamma_c$, the dashed line marks the $t^{-\theta}$
  power law with $\theta=1.81$. Data have been obtained by averaging
  over three realizations with system size $L=8192$ and by a further
  logarithmic-window average in time.  (b) The order parameter is multiplied by
  $t^{1.81}$ to obtain an asymptotically flat curve. (c) Subcritical
  behavior computed from simulations at system size $L=4096$. The
  dashed red line marks a slope of $2.19$.}
\label{fig:MNmap}
\end{figure}

Once the critical point is known, we can estimate the critical
exponents.  In Fig.~\ref{fig:MNmap}a-b we report the results on the
time decay of the order parameter at criticality. The critical
exponent $\theta$ is estimated 
by multiplying
$\rho(t)$ by $t^{\theta}$ and varying $\theta$ as to maximize the size
of the flat plateau.  However, it is worth stressing that, at variance with
the case of discontinuous maps, here finite-size effects are more severe and
numerical artifacts
\footnote{Contrary to what reported in
  Ref.\cite{Cencini2008}, we verified that the occurrence of spurious
  saturation effects of the order parameter close to the
  critical coupling are induced by the employed numerical precision
  and not by long time correlations.} may be present.  In particular,
the asymptotic power law decay sets in at later times with respect to
discontinuous maps. Moreover, the critical exponent is larger than the
corresponding DP value, so that lack of statistics tends to plague late time
data. Therefore, it is necessary to explore large lattice sizes to
obtain reasonable scaling and rule out finite-size effects. We performed
numerical simulations in systems of size $L=8192$, averaging
over three independent realizations, obtaining slightly more than a
decade of convincing scaling behavior. We estimate $\theta
= 1.81(5)$.

As it is shown in Fig. \ref{fig:MNmap}c, the behavior of the saturated
order parameter $\rho_\infty$ in the subcritical regime 
($\gamma < \gamma_c$), allows us to measure the second critical
exponent. Our best estimates, in systems of linear size $L=4096$ 
give us $\beta = 2.19(9)$.

For phase transitions in the MN universality class, the dynamical
exponent $z$ has been conjectured (and confirmed by numerical
simulations in 1+1 dimensions) to coincide with the one associated to
the KPZ equation~\cite{Munoz97}. Indeed, as it will become clearer in
Sect.~\ref{sec:due}, the MN synchronization problem can be mapped in
the depinning transition of a bounded KPZ surface.  From this mapping
one deduces that at the critical point, where the interface
asymptotically depins, MN systems should exhibit the same space and
time correlations as free KPZ ones. Thus they are also characterized
by the same $z$ exponent.  By rescaling finite size data averaged over
many realizations with $\theta=1.81$ by means of relation (\ref{z}),
we obtained $z=1.55(8)$ (see Fig.~\ref{fig:3}b), which is
compatible with the known best estimates of KPZ dynamical exponent in
2+1 dimensions, $z_{KPZ}=1.607(3)$ as reported in
Ref.~\cite{Parisi2000}.

\section{\label{sec:due}Stochastic Models and Scaling Arguments}
As far as we know, the only measurement of MN critical exponents in
two spatial dimensions reported in literature~\cite{MN2} was obtained
by numerically investigating the associated Langevin equation.  In
order to obtain independent and accurate estimations of the critical
indexes, we studied two stochastic models which are
known, in 1+1 dimensions, to belong to the MN
universality class.

\subsection{Single Step model plus a hard wall (SSW)}

The MN universality class is closely related to the depinning of a KPZ
surface from a hard substrate. Indeed the MN Langevin
equation~\cite{munoz_review}
\begin{equation}
\partial_t \phi = b \phi - c \phi^2 +D \nabla^2 \phi + \phi \eta 
\label{MNLangevin}
\end{equation}
can be formally mapped, via the Cole-Hopf transformation $\psi({\bf
  r},t)=-\ln \phi({\bf r},t)$, onto the KPZ equation with negative
nonlinear term and bounded from below \cite{munoz}
\begin{equation}
\partial_t \psi = v_0 - c e^{-\psi} +D \nabla^2 \psi - D \left(\nabla \psi\right)^2 + \eta \,.
\end{equation}
Here, $\psi({\bf r},t)\geq 0$ and $\phi({\bf r},t)\in [0,1]$ are the
coarse-grained height and difference field, respectively, while
$\eta({\bf r},t)$ is a spatio-temporal Gaussian white noise, and
$D>0$. Note that the $\phi({\bf r},t)=0$ absorbing state of the
Langevin dynamics corresponds to an infinite height (i.e. a completely
depinned) interface in the bounded KPZ representation.

The above link suggests us to consider a simple solid-on-solid
stochastic deposition model, belonging to the KPZ universality class,
such as the well known single step model (SSM), which in one spatial
dimension can be exactly mapped onto the KPZ equation (see, e.g.,
Ref.~\cite{barabasi}, see also Ref.\cite{Odor} for a study in two
spatial dimensions).  Equipped with a hard substrate, the so-called
SSM-plus-wall (SSW) provides an example of MN phase transition.  In
1+1 dimension, the critical point is analytically known and very
accurate numerical estimates of MN critical exponents have been
obtained \cite{Ginelli2003b, Ginelli2005}. 
The SSW time-depinning exponent $\theta$ has also
been computed in one spatial dimension via a mean field-like approximation
in Ref. \cite{MF}.

Here, we numerically investigate the following two dimensional
generalization of the SSW model. A fluctuating interface with an
integer height field $h_t(x,y)$ is defined on a square lattice ($x,y =
1, 2, \ldots, L$) with periodic boundary conditions. The dynamics of
the interface is subjected to the following restrictions: the heights
on nearest neighbours sites cannot be equal and must differ by one,
i.e. $|h_t(x,y)-h_t(x\pm 1,y)|=|h_t(x,y)-h_t(x,y\pm 1)|=1$, moreover a
hard lower wall at height $H_t$ moving upward with velocity $V_w$ is
imposed requiring that $h_t(x,y)>H_t$, with $H_t= V_w t$. The dynamics
is asynchronous: at each sub-time step $dt=1/L^2$ a site $(x,y)$ is
chosen at random and its height is increased by two,
$h_{t+dt}(x,y)=h_t(x,y)+2$, if $h_t(x,y)$ is a local minima.  Finally,
the wall moves up by one unit every $L^2/V_w$ time steps pushing up the
interface: every site whose height is less than $H_t$ is raised by two
units. The density $\rho$ of interface sites attached to the interface
(i.e. $h_t(x,y)=H_t$) {\it immediately after} a wall movement is the
proper order parameter, while the wall velocity $V_w$ is the control
parameter.

\begin{figure}[t!]
\centering
\includegraphics[clip=true,keepaspectratio,width=.9\textwidth]{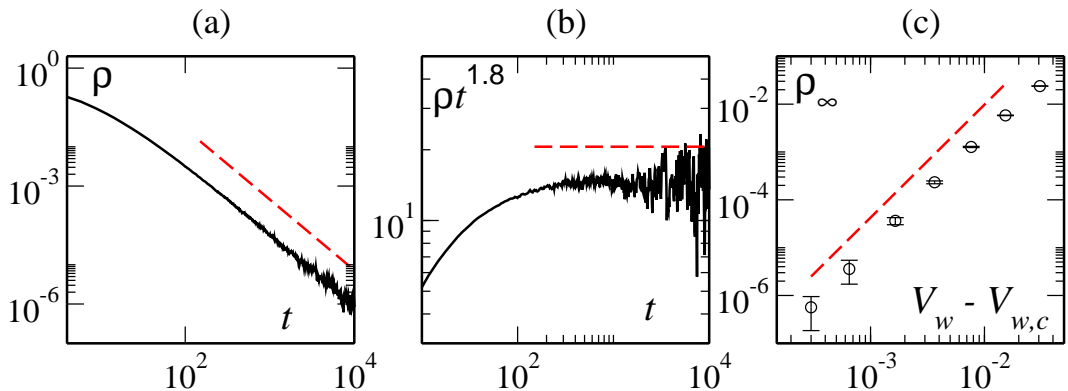}
\caption{Critical behavior of the depinning transition in the $d=2$ SSW
  model with $L=4096$.  (a) Power law decay of the order parameter at
  criticality, the dashed line marks the power law $t^{-1.8}$. Data
  are obtained by averaging three different realizations and by a
  further logarithmic-window average in time.  (b) Order parameter
  multiplied by $t^{1.8}$ to demonstrate the quality of scaling. (c) The
  subcritical behavior is characterized by a power law with exponent
  $\beta=2.36$, as indicated by the dashed line. }
\label{fig:SSW}
\end{figure}

Unfortunately, the critical velocity $V_{w,c}$ of the two dimensional
SSW is not known analytically.  Careful finite-size analysis up to
$L=4096$ has been used to locate the critical depinning velocity,
which is estimated to be $V_{w,c}=0.34135(10)$. Numerical simulations
start from a completely pinned interface:  $h_0(x,y)=0$
for $x+y$ odd and $h_0(x,y)=1$ otherwise.  Our results for the time
decay critical exponent $\theta$, shown in Fig~\ref{fig:SSW}a-b, yield
$\theta=1.80(5)$.  By slightly increasing the wall velocity above the
critical value we can  estimate the magnetization exponent,
corresponding to the behavior of the ST for subcritical couplings,
obtaining $\beta=2.36(9)$, as shown in Fig~\ref{fig:SSW}c.
Summarizing, $\theta$ is in fairly good agreement with the estimate
for the CML with skewed tent maps, and $\beta$, while still compatible
if error bars are considered, appears to be slightly larger. 
Finally, the dynamical exponent $z$ is estimated via the
finite-size scaling relation (\ref{z}) with $\theta=1.8$. A
satisfactory data collapse can be obtained with $z=1.63(5)$ (not
shown), which is compatible with the value obtained for the tent map.

It is worth concluding by remarking that in Ref.~\cite {Ginelli2005} it
has been shown that the probability distribution of the first
depinning time (i.e. the first time at which an initially flat and
pinned surface depins) {\it does not} follow a standard finite scaling
relations in the MN case. The numerical simulations of the SSW model,
however, show that other quantities of interest, as the density of
pinned sites,  follow the typical scaling of absorbing phase
transitions. This peculiarity can be probably ascribed to the 
``weakly absorbing'' nature of the MN absorbing state \cite{munoz98}.

\subsection{\label{sec:dueb} Random Multiplier model}
We now consider the Random Multiplier (RM)  defined by the dynamics~\cite{Ginelli2003}
\begin{equation}
w_{t+1}(x,y)= \left\{
\begin{array}{ll}
1, & \mbox{w.p.} \quad \alpha\tilde{w}_t(x,y) \\
\alpha\tilde{w}_t(x,y), & \mbox{w.p.} \quad 1-\alpha\tilde{w}_t(x,y)
\end{array}
\right. \;\; {\rm if} \; \tilde{w}_t(x,y) > \Delta
\label{RMmodel1}
\end{equation}
\begin{equation}
w_{t+1}(x,y)= \left\{
\begin{array}{ll}
\tilde{w}_t(x,y)/\Delta, & \mbox{w.p.} \quad \alpha \Delta \\
\alpha\tilde{w}_t(x,y), & \mbox{w.p.} \quad 1-\alpha \Delta
\end{array}
\right. \qquad\, {\rm if} \; \tilde{w}_t(x,y) \le \Delta
\label{RMmodel2}
\end{equation}
where ``w.p.'' is the shorthand notation for ``with probability'',
$w_t(x,y) \in [0:1]$ and $\tilde{w}_t(x,y) =
\nabla^2_\varepsilon w_t(x,y)$ is the discrete diffusive operator,
as in Eq.~(\ref{CML}), with $\varepsilon=4/5$, on a square lattice
($x,y = 1, 2, \ldots, L$) with periodic boundary condition.

Before discussing the results of the model, it is worth briefly
reviewing its properties and its meaning. The model
(\ref{RMmodel1})-(\ref{RMmodel2}) was originally introduced in
Ref.~\cite{Ginelli2003} to describe the synchronization error
evolution in proximity of the synchronization transition of both
continuous and discontinuous maps.  Essentially the value of parameter
$\Delta$ discriminates between discontinuous or continuous character
of the local dynamics, switching the behavior from the DP to the MN
universality class.  To better understand the origin of the model,
first consider Eq.~(\ref{CML}) equipped with the Bernoulli
map~(\ref{Bernoulli}). One can formally compute the linearized
evolution equation for the synchronization error, which reads
\begin{equation}
w_{t+1}(x,y) = \tilde{w}_t(x,y) (1-2\gamma) \partial_u F(\tilde{u}_t(x,y))\,.
\label{linearized}
\end{equation}
Such an equation holds locally for any finite synchronization error
$w_t(x,y)$ such that $\tilde{u}_t(x,y)$ and $\tilde{v}_t(x,y)$ fall on
the same branch of the Bernoulli map. In this case one simply has
$w_{t+1}(x,y)=\tilde{w}_t(x,y)\,a\,(1-2\gamma)$.  However, whenever
$\tilde{u}_t(x,y)$ and $\tilde{v}_t(x,y)$ fall on the two different
branches of the Bernoulli map, the synchronization error is typically
expanded to order $1$ values, a fact overlooked by the
linearization. This latter situation can occur with a probability
proportional to $\tilde{w}_t(x,y)$ \cite{Ginelli2003}. Setting 
$\Delta=0$, so that only Eq.~(\ref{RMmodel1}) is relevant,
reproduces the Bernoulli map dynamics with $\alpha=a(1-2\gamma)$.

On the other hand, for the skewed tent map (\ref{skewed}) --- as well
as any continuous map --- the full dynamics is well captured by
linearized dynamics. For the local map (\ref{skewed}) 
the local multiplier assumes one of the two values
$\partial_u F(a,u)=a$ or $\partial_u F(a,u)=a/(1-a)$ according to the
chaotic dynamics. By approximating the chaotic signal with randomly
chosen multipliers, the RM model with finite $\Delta$ mimics exactly this
latter situations, cfr. Eq.~(\ref{RMmodel2});  while
Eq.~(\ref{RMmodel1}) simply provides a nonlinear saturation effect.
Interestingly, in Ref.~\cite{Ginelli2003} it has been shown that there
exist a threshold $\bar{\Delta}>0$ below which the transition still
belongs to the DP class. This corresponds to the case of
almost-discontinuous piecewise linear maps on the unit interval,
characterized by a very steep branch (with slope $1/\Delta$ and width
of order $\Delta$); for a complete discussion see
Ref.~\cite{Ginelli2003} and \cite{rechtman}.

\begin{figure}[t!]
\centering
\includegraphics[clip=true,keepaspectratio,width=.9\textwidth]{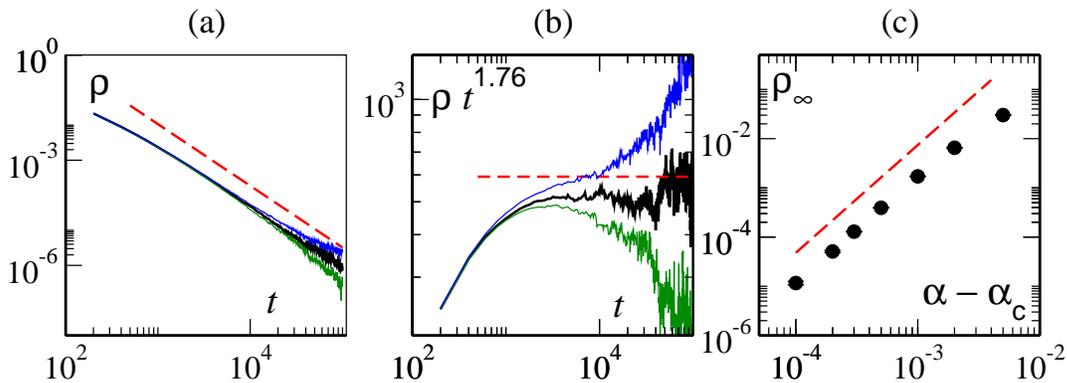}
\caption{Critical behavior of the RM model in 2+1 dimensions with size
  $L=2000$.  (a) Power law decay of the order parameter at
  criticality. Data have been obtained by averaging over $100$
  realizations.  The dashed line marks the power law $t^{-1.76}$.  (b)
  Order parameter multiplied by $t^{1.76}$; notice the asymptotic
  plateau at large times. (c) Subcritical behavior computed from up to
  $400$ independent realizations, from which $\beta$ is estimated to
  be $2.18$ (dashed line).}
\label{fig:RM}
\end{figure}

In the following, being interested to the MN transition, we
investigate the RM model for $\Delta=0.2$, which is
a sufficiently large parameter value to drive the
system into its ``linear'' regime.  The parameter $\alpha$ is the
control parameter of the transition; indeed, from
Eq.~(\ref{linearized}) it follows that $\alpha$ is essentially
equivalent to $(1-2\gamma) \partial_u F(a,u)$ so that to increase the
coupling $\gamma$ in the deterministic model amounts to decrease
$\alpha$ in the stochastic one. The synchronized regime is thus
obtained for $\alpha$ {\it smaller} than the critical value $\alpha_c$.

The critical coupling $\alpha_c$ implies a zero transverse Lyapunov exponent;
it can be estimated evaluating the value
of $\alpha$ for which $\langle \ln w_t\rangle$ grows at most logarithmically in the linear regime 
\cite{Ginelli2003}.  
This estimation is in agreement with the usual analysis performed to
evaluate the critical point from the scaling of the order parameter in
time (Fig.~\ref{fig:RM}a). In particular, by considering lattices of
linear sizes $L=2\times 10^3$ and $L=4\times 10^3$ the best scaling of
$\rho(t)$ is observed at $\alpha_c = 0.529825(5)$, in agreement with
the Lyapunov estimate which gives $\alpha_c = 0.52981(1)$. Moreover,
the critical exponent is found to be $\theta = 1.76(5)$, as shown in
Fig.~\ref{fig:RM}a-b.  For $\alpha$ slightly above $\alpha_c$, from
the saturated value of the asymptotic density we also find
$\beta=2.18(8)$ (see Fig. \ref{fig:RM}c). Finally, finite size scaling
via Eq.~(\ref{z}) gives $z=1.7(1)$ (not shown).

\subsection{Scaling arguments and comparison among the critical
  exponents\label{sec:z}}
This subsection is devoted to a discussion of the results obtained so
far for the MN universality class.  Our three independent estimates of
the critical exponents for the MN class in 2+1 dimensions are reported
in Table~\ref{tableMN} 
For $\theta$ and $\beta$, ST in CMLs, the SSW and RM model are in
remarkable agreement, the exponents are essentially coincident within
error bars. In particular, $\theta$ also coincides with the early
estimates of Ref.~\cite{MN2}, while a remarkable difference (by a
factor two) is observed for $\beta$. This is most likely due to the
preliminary nature of the $d=2$ Langevin dynamics simulations \cite{private}.
It should be noted that typically a fairly small
lack of accuracy in the estimation of the critical point can result in
a quite large error in evaluating the exponent $\beta$.  Also the
dynamical exponent $z$ measured from finite size scaling essentially
agrees with the expected KPZ value, see Table~\ref{tableMN}.

\begin{table}[hbt!]
\centering
\begin{tabular}{||c|c|c|c|c||} \hline 
  & $\theta$ & $\beta$ & z & $\nu_\parallel$ \\\hline \hline
  CML (skewed tent maps)  &   1.81(5)  &  2.19(9) & 1.55(8) &  1.21(8) \\
  RM   &   1.76(5)  &  2.18(8)   & 1.63(5) & 1.25(8)   \\
  SSW   &   1.80(4)  & 2.36(9)   & 1.7(1)   &  1.31(4)   \\
  KPZ \cite{Parisi2000} and scaling relation \cite{Munoz97} &  & &1.607(3) & 1.32(1) \\
  \hline
\end{tabular}
\caption{Critical exponents for CMLs of
  skewed tent maps, random multipliers model (RM) and
  single step model with a wall (SSW).
  Also 
  the values that can be deduced from scaling arguments and the
  KPZ numerical estimates are reported.}
\label{tableMN} 
\end{table}

Exploiting the relation with free (without walls) KPZ scaling, it has
also been conjectured that \cite{Munoz97}
\begin{equation}
\nu_\perp=\frac{1}{2(z-1)}
\label{tcorr}
\end{equation}
which, from $z_{KPZ}=1.607(3)$ as obtained in Ref.~\cite{Parisi2000},
implies $\nu_\perp=0.824(5)$ and $\nu_\parallel=z \nu_\perp
=1.32(1)$. This latter exponent can be compared with the ratio
$\nu_\parallel=\beta/\theta$ as obtained for the models here analyzed:
as reported in Table \ref{tableMN}, there is an agreement within error
bars with the KPZ estimate, as far as the RM and SSW are concerned. On
the other hand, $\nu_\parallel$ is slightly underestimated when CMLs
of skewed tent maps are considered.  
Altogether, KPZ estimates and the derived (via Eq. (\ref{tcorr})
temporal correlation exponent, well agree
with our results
in $2+1$ dimensions for both the dynamical exponent $z$ and the ratio
$\beta/\theta$.

\section{\label{sec:tre}Synchronization in the presence of mismatch}

As discussed in the introduction, in typical experimental settings it
is impossible to produce two exact replicas of the same
system. Systematic errors, slight differences in the preparation of
the system, or inhomogeneous external influence should always be taken
into account. Such small differences can be mimicked at the level of
the CMLs model as a quenched random mismatch between the dynamical
parameters of two otherwise identical systems (this idea was
introduced in Ref.~\cite{PG91} in the context of low dimensional
maps). In practice, we consider the following model, written in
generic spatial dimension $d$:
\begin{eqnarray}
u^{(1)}_{t+1}(\bm r) &=& (1-\gamma) F(a_1({\bm r}),\tilde{u}^{(1)}_t(\bm r))
+\gamma F(a_2({\bm r}),\tilde{u}^{(2)}_t(\bm r))\nonumber\\
u^{(2)}_{t+1}(\bm r) &=& (1-\gamma) F(a_2({\bm r}),\tilde{u}^{(2)}_t(\bm r))
+\gamma F(a_1({\bm r}),\tilde{u}^{(1)}_t(\bm r))
\label{eq:mismatch}
\end{eqnarray}
where ${\bm r}$ is a $d$ dimensional vector and 
\begin{equation}
\tilde{u}^{(i)}(\bm r)=(1-\varepsilon) u^{(i)}(\bm r)+\frac{\varepsilon}{2d}
\sum_{\bm r' \in NN\{\bm r\}} u^{(i)}(\bm r')\,,
\end{equation}
where the sum runs over the nearest neighbours of ${\bm r}$, denoted
as $ NN\{\bm r\}$.  The local map $F(a,u)$ depends both on the
dynamical variable $u \equiv \tilde{u}^{(i)}_t({\bm r})$ and on the
quenched parameter $a \equiv a_i({\bm r}$) (where $i=1,2$ labels the
replica). Periodic boundary conditions are considered as usual in a
$d$-cube of linear dimension $L$.  We consider two cases, the skewed
tent map defined in Eq.~(\ref{skewed}) and the Bernoulli one
(Eq.~(\ref{Bernoulli})).  Without loss of generality, we can write the
map parameter $a$ as
\begin{equation}
a_i({\bm r}) =a_0 + \omega_i({\bm r}), 
\end{equation}
with $\omega_i({\bm r})$ being a quenched random variable uniformly
distributed in $[-h,h]$.

Let us now consider the linearized dynamics for the synchronization error 
$w_t(\bm r)=u^{(1)}_t(\bm r)-u^{(2)}_t(\bm r)$,
which reads as (see also \cite{PG91})
\begin{equation}
\strut\hspace{-2cm} w_{t+1}(\bm r) = 
\tilde{w}_{t}(\bm r) (1-2\gamma) \partial_u F(a_1({\bm r}),\tilde{u}^{(1)}_t(\bm r))
+ (1-2\gamma) \delta a(\bm r) \partial_a F(a_1({\bm r}),\tilde{u}^{(1)}_t(\bm r))\,,
\label{eq:mismatch2}
\end{equation}
where $\delta a(\bm r) \equiv a_1(\bm r)-a_2(\bm r)$ is the {\it
  parameter mismatch}.  Obviously, $||\delta a(\bm r)|| \sim h$
(regardless of the chosen norm).  The first term on the r.h.s. of
Eq.~(\ref{eq:mismatch2}) --- i.e. the fluctuating ``field derivative''
$(1-2\gamma)\partial_u F(a,u)$ --- leads as usual to a stochastic term
{\it proportional} to the amplitude of the synchronization error
itself.  While the second term --- i.e. the ``parameter derivative''
$(1-2\gamma)\partial_a F(a,u)$ ---, also typically fluctuates
according to local dynamics, but it depends only on the parameter
mismatch amplitude and thus acts as an effective ``external field''
which locally prevent the complete synchronization of two replicas. At
a field theoretical level, this suggests to describe the parameter
mismatch as an external driving field with amplitude $h$.  For
instance, considering the MN class, one can add an additive
Gaussian\footnote{Note that $\langle \zeta\rangle$ can be always set
  to zero by an appropriate transformation of the field, $\phi \to \phi
  + q$.}  white noise $\zeta$ to the Langevin equation
(\ref{MNLangevin})
\begin{equation}
\partial_t \phi = b \phi - c \phi^2 +D \nabla^2 \phi + \phi \eta
+ h \zeta \,.
\label{MNLangevin2}
\end{equation}
Implicit in the above formulation is the assumption that the noise
terms $\eta$ and $\zeta$ are completely decorrelated. Of course, the
statistical independence between the second and the first term in the
r.h.s. of Eq.~(\ref{eq:mismatch2}) has to be tested a posteriori.
Unfortunately, we are not able to solve analytically 
Eq.~(\ref{MNLangevin2}). 
While efficient numerical methods \cite{Chate2005} are known 
to directly simulate multiplicative noise Langevin equations,   
we choose instead to focus on microscopic models. We consider the
RM model (\ref{RMmodel3}-\ref{RMmodel4}).  The external field then can
be described as an extra additive white noise $\chi(\bm r)$ uniformly
distributed in $[0,h]$, so that
\begin{equation}
w_{t+1}({\bm r})= \left\{
\begin{array}{ll}
1 + \chi_t({\bm r}), & \mbox{w.p.} \quad \alpha\tilde{w}_t({\bm r}) \\
\alpha\tilde{w}_t({\bm r}) + \chi_t({\bm r}), & \mbox{w.p.} \quad 1-\alpha\tilde{w}_t({\bm r})
\end{array}
\right. \;\; {\rm if} \; \tilde{w}_t({\bm r}) > \Delta
\label{RMmodel3}
\end{equation}
\begin{equation}
w_{t+1}({\bm r})= \left\{
\begin{array}{ll}
\tilde{w}_t({\bm r})/\Delta+ \chi_t({\bm r}) , & \mbox{w.p.} \quad \alpha \Delta \\
\alpha\tilde{w}_t({\bm r}) + \chi_t({\bm r}), & \mbox{w.p.} \quad 1-\alpha \Delta
\end{array}
\right. \qquad\, {\rm if} \; \tilde{w}_t({\bm r}) \le \Delta
\label{RMmodel4}
\end{equation}
Indeed, the additive noise term $\chi({\bm r})$ creates local activity
with an amplitude proportional to $h$, thus preventing complete
synchronization between the two replicas. \\
In the following we compare direct numerical simulations of the
modified RM model (\ref{RMmodel3}-\ref{RMmodel4}) with the ones of the
coupled CMLs (\ref{eq:mismatch}) with skewed tent maps with a quenched
mismatch.  We expect the asymptotic order parameter $\rho_\infty$ to
saturate to an $h$-dependent value, that   at criticality
should scale as  
\begin{equation}
\rho_\infty \sim h^\kappa
\end{equation}
where $\kappa$ is a new critical index (associated to the external
field in the field theoretical treatment).

Choosing $\langle a_i({\bm r})\rangle =a_0=2.2$, the critical point
estimated in Sec.~\ref{sec:uno} ($\gamma_c=0.13176$) remains unchanged
for $h>0$.  Equally, we set $\Delta=0.2$ for the RM model so that
$\alpha_c=0.52981$.

\begin{figure}[t!]
\centering
\includegraphics[clip=true,keepaspectratio,width=0.9\textwidth]{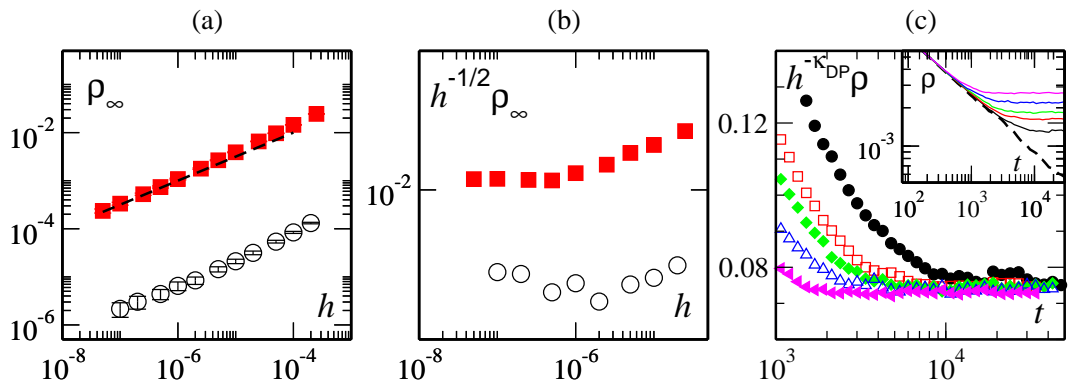}
\caption{Critical dependence on the mismatch in CMLs in 2+1
  dimensions.  (a) Mismatch amplitude dependence of the saturated
  order parameter at criticality for skewed maps CMLs (black circles)
  and RM with an additive noise (red squares). The dashed red line
  marks the $h^{1/2}$ power law. (b) Same data as (a) rescaled by a
  factor $h^{-1/2}$ to highlight critical behavior.  System size is
  $L=2048$.  (c) For Bernoulli maps we check the DP expectation
  $\kappa_{DP}=\beta/\sigma$ by plotting $\rho(t) h^{-\kappa_{DP}}$
  (see text): notice the collapse on a single plateau at large
  times. From top to bottom $h=10^{-7}$, $2.5\times 10^{-7}$,
  $6.25\times 10^{-7}$, $10^{-6}$, $2\times 10^{-6}$. System sizes is
  $L=1024$ and data have been averaged over $50$ initial
  conditions. Inset: non rescaled data compared with the $h=0$ behavior
  (dashed black line).  }
\label{fig:MNmism}
\end{figure}
Starting from random initial conditions, we measure $\rho_\infty$ as a
function of $h$ in both one and two spatial dimensions.  The results
in two spatial dimensions, reported in Fig.~\ref{fig:MNmism}a-b, show
a remarkable agreement between CMLs and the modified RM model, thus
confirming that the correlations between the additive and the
multiplicative terms in Eq.~(\ref{eq:mismatch2}) are irrelevant.  In
particular, we found $\kappa=0.50(5)$ for the RM and $\kappa=0.46(7)$
for the CMLs. In one spatial dimension (not shown) we have
$\kappa=0.21(2)$ (RM) and $\kappa=0.23(2)$ (CMLs). This is in rough
agreement with the $d=1$ early estimates obtained by direct simulation
of the Langevin equation (\ref{MNLangevin2}) in Ref.~\cite{Munoz97}.
In Ref.~\cite{Munoz96} a zero dimensional version of
Eq.~(\ref{MNLangevin2}) has been exactly solved via a Fokker-Planck
approach to yield a logarithmic increase with the field $h$.  While we
have not been able to put forward any analytical argument in $d>0$,
these results lead us to conjecture $\kappa = d/4$ for MN in an
external field.

We finally consider the case of Bernoulli map with $\langle a_i({\bm
  r})\rangle=a_0=2$.  Here, we can exploit the fact that the scaling
theory for Directed Percolation in an external field is well known
(see \cite{haye} and references therein), and thus test if the
dependence of $\rho_\infty$ on $h$ follows the DP prediction. In
particular, at criticality, the order parameter should saturates to an
asymptotic value $\rho_\infty(h) \sim h^\kappa$ with $\kappa = \beta
/\sigma$, where $\beta$ links to the behavior of the density close to
the critical point and $\sigma$ controls the mean size of inactive
clusters close to the critical point, and is related to the other
exponents by the formula \cite{haye}
$$
\sigma=\nu_\parallel(d/z+1-\theta)=\nu_\parallel+d\nu_\perp-\beta\,.
$$ 
By means of the best available DP numerical estimates \cite{haye}, we
have $\kappa_{DP}=0.108247(4)$ in $d=1$ and $\kappa_{DP}=0.268(4)$ in
$d=2$.  As shown in Fig.~\ref{fig:MNmism}c direct numerical
simulations of $2+1$ dimensional CMLs with mismatched Bernoulli maps
are in very good agreement with this prediction. This is also true in
$1+1$ dimensions (not shown).

\section{\label{sec:quattro}Conclusions}

The synchronization transition between spatially extended systems
exhibiting spatio-temporal chaos is a prototypical example of a
fluctuation driven phase transition induced by microscopic chaos. The
present numerical study in 2+1 dimensions has clearly confirmed that,
analogously to the 1+1 dimensional case, the synchronization
transitions can been described in the framework of continuous
out-of-equilibrium critical phenomena towards an absorbing phase. In
particular, depending on the perturbation propagation properties of
the spatio-temporal dynamics, the synchronization transition belongs
to two possible universality classes, namely Directed Percolation and
Multiplicative Noise.  The above results confirm that the ST belongs
to the DP universality class whenever the transverse Lyapunov exponent
is negative at the critical point. Differently, MN behavior sets in
when a zero transverse Lyapunov exponent characterize the critical
point. As for the latter universality class, we have produced the best
available numerical estimates of the critical exponents in two spatial
dimensions.  Furthermore, by analyzing different deterministic and
stochastic models, we have been able to confirm universality in the MN
class and its link with deposition processes.

We have also addressed the effect of a small difference in the
dynamics of the two replicas, an experimentally relevant question
given the  practical impossibility to produce two exactly identical
systems in any experimental setup. By modeling this difference as a
quenched parametric mismatch in the local dynamics, we have shown that
it amounts to the action of an external field within the Langevin
description. Numerical simulations in one and two spatial dimensions
for discontinuous local maps confirm DP scaling theory, which predicts
a power law dependence of the saturated density from the external
field with an exponent $\kappa$ depending from the usual zero field
critical ones.  Finally, we obtained an estimate for this exponent
also for MN in en external field (smooth maps).

Before concluding, a comment on the behavior in higher dimensions is
in order.  In $d>2$ naive power counting in the MN Langevin
Eq.~(\ref{MNLangevin}) predicts the coexistence of two different fixed
points in the Renormalization Group flow, a mean field one acting at
small but finite noise amplitude and a strong coupling one at large
noise amplitudes \cite{Munoz96}. 
Estimates obtained from numerical simulations of Eq.~(\ref{MNLangevin}) for 
MN in 3+1 dimensions can be found in Ref.~\cite{Munoz99}.

We conclude expressing our hope that these results will stimulate
experimental studies on the synchronization of extended
systems exhibiting spatio-temporal chaos.

\ack{We acknowledge M.A. Mu\~noz, A. Pikovsky, A. Politi, H. Chat\'e,
  I. Dornic and K.A. Takeuchi for fruitful discussions. MC and AT were
  partially supported by the Italian project ``Dinamiche cooperative
  in strutture quasi uni-dimensionali'' N. 827 within the CNR
  programme ``Ricerca spontanea a tema libero''.}


\section*{References}


\begin{thebibliography}{99}

\bibitem{PikoBook} Pikovsky A S, Rosenblum M and Kurths J,
\textit{Synchronization:  A Universal Concept in Nonlinear Sciences},
2001 (Cambridge: Cambridge University Press) 

\bibitem{baroni} Baroni L, Livi R, and Torcini A, in {\it Dynamical
  Systems:from Cristal to Chaos},  Gambaudo J M, Hubert P,
   Tisseur P and Vaienti S eds., 2000 (Singapore: World Scientific) p. 23;
  Baroni L, Livi R and Torcini A, 2001 \textit{Phys. Rev. E} \textbf{63} 036226

\bibitem{PA02} Ahlers V and Pikovsky A S,  2002  \PRL \textbf{88} 254101

\bibitem{PPK} Pikovsky A S and Kurths J, 1994 \PRL \textbf{49} 898;
  Pikovsky A S and Politi A, 1998 \textit{Nonlinearity} \textbf{11}
  1049

\bibitem{haye} Hinrichsen H,  2000 \textit{Adv. Phys.} \textbf{49} 815

\bibitem{PG91} Pikovsky A S and Grassberger P, 1991 \JPA \textbf{24} 4587 

\bibitem{Ginelli2003} Ginelli F,  Livi R, 
Politi A and Torcini A, 2003 \textit{Phys. Rev. E} \textbf{67} 046217

\bibitem{Munoz2003} Mu\~noz M A, de los Santos F and Achahbar A,
  2003 \textit{Braz. J. Phys.} \textbf{33} 443.

\bibitem{munoz_review} Mu\~noz M A, in
\textit{Advances in Condensed Matter and  Statistical Mechanics},
Korutcheva  E, \textit{et al.}  eds., 2004 (New York: Nova Science Publishers)

\bibitem{cencini_torcini} Torcini A, Grassberger P and Politi A, 1995
  \JPA \textbf{28} 4533; Cencini M and Torcini A, 2001
  \textit{Phys. Rev. E} \textbf{63} 056201


\bibitem{SC} Politi A, Livi R, Oppo G L and Kapral R, 1993
  \textit{Europhys. Lett.}  \textbf{22} 571; Politi A and Torcini A,
  2009 arXiv:0902.2545

\bibitem{rechtman} Bagnoli F, and Rechtman R, 2006 \textit{Phys. Rev. E}
  \textbf{73}  026202

\bibitem{KPZ}  Kardar M,  Parisi G, and  Zhang Y C, 1986 \PRL  \textbf{56} 889


\bibitem{Ginelli2003b} Ginelli F, Ahlers V, Livi R, Mukamel D,
  Pikovsky A S, Politi A and Torcini A, 2003 \textit{Phys. Rev. E}
  \textbf{68} 065102

\bibitem{munoz}  Mu\~noz M A and  Pastor-Satorras R,  2003 \PRL \textbf{90} 204101 

\bibitem{cecconi} Bagnoli F, Baroni L, and Palmerini P, 1999 \textit{Phys. Rev. E}
  \textbf{59} 409; Bagnoli F and Cecconi F, 2001 \textit{Phys. Lett. A}
  \textbf{282} 9

\bibitem{droz} Droz M and Lipowski A, 2003 \textit{Phys. Rev. E}
  \textbf{67} 056204; 2003 {\it ibidem} \textbf{68} 056119

\bibitem{gade} Gade P M and Hu C-K, 2006 \textit{Phys. Rev. E} \textbf{73}
  036212

\bibitem{szendro09} Szendro I G, Rodriguez M A, and L\'opez J M, 2009
  \textit{Europhys. Lett.} \textbf{86} 20008

\bibitem{Bagnoli99} Bagnoli, F and Rechtman R, 1999 \textit{Phys. Rev. E}
  \textbf{59}  R1307

\bibitem{Morvan} Rouquier J B, and Morvan M, 2009 \textit{J. Cell. Auto.}
{\bf 4} 55

\bibitem{Tessone2007} Tessone C J,  Cencini M and Torcini A, 2006 \PRL \textbf{97} 224101

\bibitem{Cencini2008} Cencini M, Tessone C J and Torcini A, 2008
  \textit{Chaos} \textbf{18} 037125

\bibitem{mohanty} Mohanty P K, 2004 \textit{Phys. Rev. E} \textbf{70}
  045202(R)

\bibitem{szendro} Szendro I G and L\'opez J M, 2005 \textit{Phys. Rev. E}
  \textbf{71} 055203(R)


\bibitem{Ginelli2005}  Kissinger T, Kotowicz A,  Kurz O,  Ginelli F and  Hinrichsen H,  2005 JSTAT P06002

\bibitem{MN2} Genovese W, Mu\~noz M A and Sancho J M, 1998
  \textit{Phys. Rev. E} \textbf{57} R2495



\bibitem{Munoz96}  Grinstein G, Mu\~noz  M A and Tu Y, 1996 \PRL  
\textbf{76} 4376


\bibitem{Kuramoto}  Kuramoto Y, \textit{Chemical Oscillations, 
Waves and Turbulence}, 1984 (Berlin: Springer-Verlag)

\bibitem{exper} Hildebrand M, Cui J, Mihaliuk E, Wang J and Showalter
  K, 2003  \textit{Phys. Rev. E} \textbf{68} 026205

\bibitem{Chate}  Takeuchi K A, Kuroda M,  Chat\'e H and Sano M,
 2007 \PRL \textbf{99} 234503; and preprint arXiv:0907.4297v1


\bibitem{tedeschi} Rupp P, Richter R, and Rehberg I, 2003 \textit{Phys. Rev. E} \textbf{67} 036209

\bibitem{munoz98} Mu\~noz M A, 1998. \textit{Phys. Rev. E} \textbf{57},
  1377



\bibitem{PT94}
Politi A and Torcini A, 1994 \textit{Europhys. Lett.} \textbf{28} 545



\bibitem{Munoz97}   Tu Y, Grinstein G and  Mu\~noz M A, 1997 \PRL \textbf{78} 274

\bibitem{Parisi2000} Marinari E, Pagnani A and  Parisi G, 
2000 \JPA \textbf{33} 8181 


\bibitem{barabasi} Barab\'asi A L and Stanley H E, \textit{Fractal
    Concepts in Surface Growth}, 1995 (Cambridge: Cambridge University
  Press)

\bibitem{Odor} Odor G, Liedke B and Heinig K-H,
  2009 \textit{Phys. Rev. E} \textbf{79} 021125

\bibitem{MF} Ginelli F, and Hinrichsen H, 2004 \JPA \textbf{37} 11085

\bibitem{private} Mu\~noz M A, {\it private communication}.

\bibitem{Chate2005}  Dornic I, Chat\'e H, and Mu\~noz M A,
 2005 \PRL \textbf{94} 100601 

\bibitem{Munoz99} Genovese W and Mu\~noz M A, 1997 \textit{Phys. Rev. E}
  \textbf{60}  69
  

\end{thebibliography}
\end{document}